\documentclass[11pt]{article}
\usepackage{fullpage} 
\usepackage{setspace}
\usepackage{cite}
\usepackage{lineno,hyperref}
\hypersetup{colorlinks = true,linkcolor = blue,anchorcolor =red,citecolor = blue,filecolor = red,urlcolor = red,
            pdfauthor=author}
\modulolinenumbers[5]

\newcommand{\exactslow}{\mathtt{Integer Token Passing}}
\newcommand{\approxslow}{\mathtt{Powers Of Two Token Passing}}
\newcommand{\approxjunta}{\mathtt{Junta Election}}
\newcommand{\averaging}{\mathtt{Discrete Averaging}}
\newcommand{\powertwoaveraging}{\mathtt{Powers Of Two Averaging}}
\newcommand{\sen}{\mathtt{sen}}
\newcommand{\rec}{\mathtt{rec}}
\newcommand{\agent}{\mathtt{agent}}
\newcommand{\innput}{\mathtt{input}}
\newcommand{\outtput}{\mathtt{output}}
\newcommand{\ave}{\mathtt{average}}
\newcommand{\countt}{\mathtt{count}}
\newcommand{\codename}{\mathtt{codename}}
\newcommand{\exponent}{\mathtt{exponent}}
\newcommand{\A}{\mathtt{A}}
\newcommand{\B}{\mathtt{B}}
\newcommand{\actve}{\mathtt{active}}
\newcommand{\level}{\mathtt{level}}
\newcommand{\True}{\mathsf{True}}
\newcommand{\False}{\mathsf{False}}
\newcommand{\Not}{\mathsf{not}}
\newcommand{\upstream}{\textbf{U}}
\newcommand{\downstream}{\textbf{D}}

\usepackage[textsize=scriptsize]{todonotes} 
\usepackage{algorithm}
\usepackage{caption}
\usepackage[noend]{algpseudocode}

\newcommand{\T}{\mathsf{T}}
\newcommand{\F}{\mathsf{F}}

\newcommand{\polylog}{\mathrm{polylog}}
\newcommand{\poly}{\mathrm{poly}}

\usepackage{mathtools}
\newcommand{\bfloor}[1]{\left\lfloor#1\right\rfloor}
\newcommand{\bceil}[1]{\left\lceil#1\right\rceil}
\DeclarePairedDelimiter\ceil{\lceil}{\rceil}
\DeclarePairedDelimiter\floor{\lfloor}{\rfloor}

\usepackage{amsfonts}

\newcommand{\N}{\mathbb{N}}

\usepackage[normalem]{ulem}

\title{\Large A survey of size counting in population protocols}

\usepackage{authblk}
\author[1]{David Doty}
\affil[1]{University of California, Davis, {\tt doty@ucdavis.edu}}

\author[2]{Mahsa Eftekhari}
\affil[2]{University of California, Davis, {\tt mhseftekhari@ucdavis.edu}}

\date{}

\begin{document}

\maketitle

\begin{abstract}
The population protocol model describes a network of $n$ anonymous agents who cannot control with whom they interact. 
The agents collectively solve a computational problem through random pairwise interactions, each agent updating its state in response to seeing the state of the other agent. 
Population protocols are equivalent to the model of chemical reaction networks, describing abstract chemical reactions such as $A+B \to C+D$, when the latter is subject to the restriction that all reactions have two reactants and two products, and all rate constants are 1.
The \emph{counting} problem is that of designing a protocol so that $n$ agents, all starting in the same state, eventually converge to states where each agent encodes in its state an exact or approximate description of population size $n$.
In this survey paper, we describe recent algorithmic advances on the counting problem.
\end{abstract}

\paragraph{keywords}
Population protocols, Population size counting, Exact counting, Approximate counting

\paragraph{Acknowledgements.}
Doty and Eftekhari were supported by NSF award 1900931 and CAREER award 1844976.

\thispagestyle{empty}
\clearpage 
\pagenumbering{arabic} 

\section{Introduction}
\label{sec:intro}

A population is a network of $n$ anonymous and identical \emph{agents}, 
each holding a \emph{state} representing its entire memory. The agents communicate through a sequence of randomly chosen pairwise interactions. 
In an interaction, the scheduler selects two different agents uniformly at random. 
Each observes the state of the other and updates its own according to the transition function defined by a protocol. 
A protocol is designed to perform a common task, e.g., leader election: selecting exactly one agent as leader over the population. 
Following the protocol drives the population from a 
valid initial configuration to a desired configuration (for example, from all agents being leaders to only one).\footnote{
    For the problems of leader election and population size computation, all agents start in the same state in a valid initial configuration.  However, 
    in some settings, the initial configuration is assumed already to contain a unique leader~\cite{AAE08, michail2018fast, CheDotSolNaCo}.
}

A protocol is defined by \emph{transitions} 
that describe, given a pair of input states of the agents that interact, how the agents should update their memory.
For example, a simple leader election transition is $(L, L) \to (L, F)$ with all agents starting in state $L$.
This protocol \emph{stabilizes}, 
meaning that with probability 1, it reaches a configuration that is both correct 
and \emph{stable}: every subsequently reachable configuration is also correct.
In the original model of population protocols~\cite{AADFP06}, the states and transitions are constant with respect to the population size $n$. 
However, 
recent studies use a variant of the model: allowing the number of states and transitions to grow with $n$.
One motivation to study population protocols with 
$\omega(1)$ states is the existence of impossibility results 
showing that no constant-state protocol can stabilize in sublinear time with probability $1$ for problems such as leader election~\cite{doty2018stable}, majority~\cite{alistarh2017time}, or 
computation of more general predicates and integer-valued functions~\cite{belleville2017time}.\footnote{
    It is conventional in population protocols to measure the memory usage by counting the total number $s(n)$ of states agents can store in population size $n$, instead of the number of bits required to represent these states, which is about $\log s(n)$.
}
The recent algorithmic advances using non-constant states~\cite{AG15, berenbrink2018simpleLE, GS18, gkasieniec2018almost, alistarh2017time, alistarh2018space, berenbrink18majority, bennun20majority, sudo2018logarithmic, bilke2017brief, MocquardAABS2015, doty2021majority, berenbrink_2020_optimalLE, Berenbrink2020majority_tradeoff}, 
lead to time- and space-optimal solutions for leader election~\cite{berenbrink_2020_optimalLE} and majority~\cite{doty2021majority} problems. 
However, most of these solutions~\cite{AG15, berenbrink2018simpleLE, Berenbrink2020majority_tradeoff,  gkasieniec2018almost,alistarh2017time, alistarh2018space, doty2021majority, berenbrink18majority, bennun20majority, berenbrink_2020_optimalLE, sudo2018logarithmic, MocquardAABS2015,bilke2017brief} propose a nonuniform protocol.
Rather than being a single set of transitions, a nonuniform protocol
represents a \emph{family} of protocols, 
where the set of transition rules (i.e., the protocol) used is allowed to depend on the population size $n$. 

The typical way that nonuniformity appears in a protocol is that the agents have an estimate of $n$
(for example the value $\ceil{\log n}$) appearing in the transitions.
When expressed in pseudocode, this is often realized by a ``hard-coded'' constant $\ceil{\log n}$ to which the code has access 
(i.e., each agent receives the value $\ceil{\log n}$ as ``advice'').
However, in measuring the state complexity of the protocol,
the space required to store this value does not count against the memory usage.
Note that this concern is relevant because we count memory complexity by counting the number of states, rather than the number of bits necessary to represent the state. Adding a field with $O(\log n)$ different values does not asymptotically change the bit usage, but it does asymptotically increase the number of states.
As an example of a nonuniform protocol using its estimate of $\log n$, 
consider the following nonuniform leaderless phase clock rules: each agent, independently, counts its number of interactions and uses it as a timer by comparing to $12 \ln n$. In this protocol, each agents increment the value in its $\countt$ field until it reaches a threshold dependent on $\ln n$ (note that ``$-$'' stands for an arbitrary value in $\N$ for $\countt$):
\begin{equation}\label{leaderless-phase-clock}
\left\{
\begin{array}{@{\,}r@{}ll}
  (\A = \F, \countt = i) , (-, -) & \longrightarrow (\A = \F, \countt = i+1), (- ,-) 
  & \textbf{if } i < 12\ln n\\
  (\A = \F, \countt = i) , (-, -) & \longrightarrow (\A = \T, \countt = 0), (-, -) & \textbf{if } i = 12\ln n
\end{array}\right.
\end{equation}

In the above leaderless phase clock, no agent will set their $\A$ field to $\T$ before $6\ln n$ time has passed 
with probability at least $1-1/n$.\footnote{
    We can compute the error probability with a straightforward Chernoff bound application on binomial random variables.
}
However, no uniform protocol can achieve this same task:
in any uniform protocol, 
some agent will set its $\A$ field to $\T$ in constant time with high probability~\cite[Theorem 4.1]{doty2018efficient}.

\paragraph*{Uniform Computation}
In a \emph{uniform} protocol, 
by contrast, the transitions are not dependent on the population size $n$, i.e., agents lack any knowledge of $n$. 



The original $O(1)$-state model~\cite{AADFP06, AAE06, AAE08} is uniform, since there is a single transition function for all population sizes.
However
, uniform protocols are not required to have constant states. 
For example, starting from $n$ agents in state $L_1$,
the protocol defined by transitions $L_i,L_j \to L_{i+j},F$ for all natural numbers $i, j$, in a population of size $n$, can produce all values of $L_i$ for $i$ between $1$ and $n$.
However, note that the (infinitely many) transitions are ``uniformly specified'': no transition makes reference to an estimate of $n$.
(This is formalized 
by requiring the transition function to be computable by a single Turing machine~\cite{CMNPS11, DEMST18}.)

\subsection{Definitions and notation}
\label{subsec:defs}

To measure a protocol's computation time, we consider the expected number of interactions, starting from the initial configuration to reach a desired configuration. 
Since we would like to model that many interactions can happen in parallel, with $O(1)$ interactions per agent per unit of time, 
we define $n$ interactions as one unit of time. 
This definition coincides with the time defined in the standard Gillespie kinetic model for chemical reaction networks~\cite{Gillespie77}, of which population protocols are a special case describing $n$ molecules reacting in a volume proportional to $n$.

We say that a protocol \emph{stably} solves a problem if the agents eventually reach a correct configuration with probability $1$, and no subsequent interactions can move the agents to an incorrect configuration; i.e., the configuration is \emph{stable}. 


In this paper, we use the term ``with high probability'' (or w.h.p.) to refer the probability of at least $1-1/n$. However, the standard definition of high probability refers to probability of at least $1-1/n^c$ for some constant $c > 0$, where $c$ can be made arbitrarily large by adjusting appropriate parameters in the algorithm.

Since in each interaction, the scheduler picks an \emph{ordered} pair of agents to interact, we denote these agents in the pseudocode as receiver ($\rec$) and sender ($\sen$) respectively.
In other words, unlike many models of distributed computing, population protocols typically are defined to be able to break symmetry ``for free''.\footnote{There are examples of interesting protocols using only symmetric transitions, and under certain circumstances, asymmetric protocols can be simulated by symmetric ones~\cite{bournez2013population}.}

\subsection{Population size counting}
Population size counting is the problem of computing the number of agents in a population protocol.
Both exact~\cite{DEMST18, BerenbrinkKR19} (computing $n$) and 
approximate counting~\cite{BerenbrinkKR19, doty2018efficient, alistarh2017time, michail2018fast} (computing $\ceil{\log n}$ or $\floor{\log n}$, which gives $2^{\ceil{\log n}}$  
or $2^{\floor{\log n}}$
as a multiplicative factor-2 estimate of $n$) 
have been considered in the literature. 
Considering the size counting problem in a nonuniform model of population protocol is trivial since we can provide the agents the values of $n$ or $\ceil{\log n}$ as advice. 
Thus, all cited papers solving this problem use the uniform variant of the model~\cite{DEMST18, BerenbrinkKR19, doty2018efficient, alistarh2017time, michail2018fast}.

\paragraph*{Motivation}
The recent algorithmic advances for population size counting problem provide composable building blocks that simplify the (uniform) solution of other problems:
compute an estimate of $\log n$, and use this value where a nonuniform protocol would use the hard-coded constant $\ceil{\log n}$. 
We can adopt a counting technique as a black box and 
compose it with a nonuniform protocol through a restarting scheme~\cite{GS18, doty2018efficient, BerenbrinkKR19} to obtain a uniform protocol. 
We explain the composition scheme in Section~\ref{sec:tools}.

In this survey paper, we will discuss the existing counting protocols and draw attention to their time-space tradeoff. We will cover both the exact and approximate counting problems, since, for most protocols, having an approximation of $\log n$ suffices. Tables~\ref{table:exact-counting-summary},~\ref{table:approx-counting-summary} summarize both exact and approximate counting protocols in the conventional model of population protocols (with initialized population).

\begin{table}[ht!]
{\small 
    \makebox[\textwidth]{
    \begin{tabular}{ |p{0.8cm}|p{0.8cm}|p{1.2cm}|p{3.1cm}|p{3.1cm}|p{2cm}| }
     \hline
     \multicolumn{6}{|c|}{Exact counting protocols} \\
     \hline
     ref. & sec. & prob. & time & states & comments \\
     \hline
        \cite{DEMST18}&\ref{subsec:doty-exact}& $1$ &$O(\log n \log \log n)$  & $O(n^{60})$ & stable\\
        \cite{BerenbrinkKR19}&\ref{subsec:berenbrink-exact}& $1-\frac{O(1)}{n}$ & $O(\log n)$ & $O(n \log n)$ & - \\
        \cite{BerenbrinkKR19}&\ref{subsec:berenbrink-exact}& $1$ & $O(\log n)$ & $O(n \log n \log \log n)$ &
        stable\\
     \hline
    \end{tabular}
    }
}
\caption{Summary of existing protocols for the exact counting problem. Note that ``stable'' means correct with probability 1. For all the stable protocols, the stated time bounds the stated time bounds are proven both with high probability and in expectation. However, the state complexity for all the protocols is correct with high probability. We also mention the correctness probability for each protocol under the ``prob.'' column. We also discuss counting in population protocols with constant message size and in the self-stabilizing model in Sections~\ref{subsec:constant-message},~\ref{subsec:self-stabilizing} respectively.
}
\label{table:exact-counting-summary}
\end{table}

\begin{table}[ht]
{\small 
    \makebox[\textwidth]{
    \begin{tabular}{ |p{1.5cm}|p{0.35cm}|p{3.5cm}|p{1.6cm}|p{2.4cm}|p{2.6cm}|p{1.7cm}| }
    
     \hline
     \multicolumn{7}{|c|}{Approximate counting protocols} \\
     \hline
     ref. & sec. & output value (range) & prob. & time & states & comments \\
     \hline
        \cite{doty2021majority, BerenbrinkKR19}& \ref{subsec:approx-slow}& $\floor{\log n}$ & $1$ & $O(n\log n)$ & $O(\log^2 n)$ & \footnotesize stable \\
        \cite{alistarh2017time}& \ref{subsec:n-geometrics}& $[\frac{1}{2}\log n, 9\log n]$ & $1-\frac{O(1)}{n^3}$ & $O(\log n)$ & $O(\log^9 n)$ & \footnotesize deterministic \\
        \cite{alistarh2017time, doty2018efficient, sudo2018logarithmic}& \ref{subsec:n-geometrics}&$[\log n- \log \ln n, 2\log n]$ & $1-\frac{O(1)}{n}$ & $O(\log n)$ & $O(\log^2 n)$ & - \\
        \cite{GS18, Berenbrink2020majority_tradeoff}&\ref{subsec:junta-election}& $[\frac{\log n}{16}, 256 \log n]$ & $1-\frac{O(1)}{n}$ & $O(\log n)$ & $O(\log \log n)$ & \footnotesize deterministic\\
        \cite{doty2018efficient}&\ref{subsec:doty-approx}&$[\log n-5.7, \log n+5.7]$& $1-\frac{O(1)}{n}$ 
        & $O(\log^2 n)$ & $O(\log^4 n)$ & - \\
         \cite{BerenbrinkKR19}&\ref{subsec:berenbrink-approx}&$\floor{\log n}$ or $\ceil{\log n}$ &$1-\frac{O(\log n)}{n^2}$ &  $O(\log^2 n)$ & $O(\log n \log \log n)$ & - \\
        \cite{BerenbrinkKR19}&\ref{subsec:berenbrink-approx}& $\floor{\log n}$ or $\ceil{\log n}$ &$1$ & $O(\log^2 n)$& $O(\log^2 n\log \log n)$  &  stable\\
     \hline
     
    \end{tabular}
    }
}
\caption{Summary of existing leaderless protocols for the approximate counting problem. The approximation factor of each protocol is implied under the ``output value''. The columns follow the same convention as Table~\ref{table:exact-counting-summary}. One leader-driven protocol is discussed in Section~\ref{subsec:michail-approx}.}
\label{table:approx-counting-summary}
\end{table}

\section{Prerequisite: Fast averaging protocol}\label{sec:prereq-mocquard-average}

The averaging technique discussed in this section does not solve the counting problem but is used in the subsequently discussed counting protocols in Sections~\ref{subsec:doty-exact}, \ref{subsec:berenbrink-exact}, and~\ref{subsec:berenbrink-approx}.

The averaging technique, also known as 
randomized load balancing~\cite{berenbrink2019loadbalancing, berenbrink2019tight,sauerwald2012tight}, was first introduced in population protocols in~\cite{alistarh2015majority} to solve the exact majority problem.

Each agent's state is an integer; for intuition, assume the integers represent a ``load''\footnote{In the rest of the paper, whenever the load values are nonnegative, we use ``tokens'' instead to present a more intuitive explanation of the protocols.} that each agent holds. The averaging rules allow each selected pair of agents to exchange loads to balance (as best they can) their values, e.g., $(2,11) \to (6,7)$. This
leads the population to a configuration in which all agents have almost equal values:
concretely, if the total load among the population is $m$, then after stabilization, each agent holds either the value $\floor{m/n}$ or $\ceil{m/n}$.
Stabilization can take $\Theta(n)$ time in the worst case, but it takes only $O(\log n)$ time for all agents to hold \emph{three} consecutive values (two of which are $\floor{m/n}$ or $\ceil{m/n}$)~\cite{berenbrink2019loadbalancing, MocquardAS2016}.
The averaging technique has been crucial in several polylogarithmic-time protocols for problems such as population size counting~\cite{DEMST18, BerenbrinkKR19} and majority related problems~\cite{MocquardAS2016,berenbrink18majority,alistarh2015majority, MocquardAABS2015}, and its time complexity has been tightly analyzed~\cite{MocquardAABS2015, berenbrink2019loadbalancing, berenbrink2019tight, mocquard2019tight}.

Notably, Mocquard, Anceaume, Aspnes, Busnel, and Sericola~\cite{MocquardAABS2015} used the averaging technique to solve a generalization of the exact majority problem.
Considering a population with $n_a$, $n_b$ initialized agents in states $\A$ and $\B$, i.e., $n_a + n_b = n$, the authors designed an averaging-based protocol that counts the \emph{exact} difference between the number of agents in the $\A$ and $\B$ (computing the value of $n_a-n_b$).

In this protocol, the $\A$ and $\B$ agents start with $+m$ and $-m$ values respectively, where $m$ is a large integer with respect to the population size $n$. 
Thus the population as a whole starts with
a total of $m(n_a) - m(n_b)$ load. 
The protocol is designed to almost equally distribute the load among the 
agents while \emph{preserving the total sum}. 
In this protocol, the agents update their state according to the ``discrete averaging'' rule described in Protocol~\ref{protocol:avg-one-way}. 

\begin{algorithm}[ht]
	\floatname{algorithm}{Protocol}
	\caption{$\averaging(\rec, \sen)$\\
	    \textbf{Initialization: }\\
	    if $\agent.\innput = \A$: $\agent.\ave \gets m$\\
	    if $\agent.\innput = \B$: $\agent.\ave \gets -m$
	}
	\label{protocol:avg-one-way}
	\begin{algorithmic}[1]
		\State {$\rec.\ave, \sen.\ave \gets
		\bceil{  \frac{\rec.\ave + \sen.\ave}{2} } ,
		\bfloor{ \frac{\rec.\ave + \sen.\ave}{2} }$
		}
		\State {$\rec.\outtput \gets \bfloor{\frac{n \times \rec.\ave}{m}+ \frac{1}{2}}$}
	\end{algorithmic}
\end{algorithm}

Initializing Protocol~\ref{protocol:avg-one-way} with $n_a$, $n_b$ agents in states $\A$ and $\B$,
it is shown that the agents'  $\ave$ value converges to $\frac{(n_a - n_b)m}{n}$ quickly. In fact, the authors of~\cite{MocquardAABS2015} proved for $m = \bceil{\frac{\sqrt{2}n^{3/2}}{\sqrt{\delta}}}$, after $O(n \log n)$ interactions with probability $1-\delta$, the $\outtput$ field of agents will be equal to $n_a - n_b$. (e.g. to achieve a high probability result, one can set $\delta = 1/n$ and start the protocol with $m = O(n^2)$).


The protocol given in~\cite{MocquardAABS2015} is nonuniform; it is assumed that the population size is known in advance.
Crucially, the protocol requires all agents to store the exact value of $n$ in their memory to compute the $\outtput$.
In a separate paper~\cite{MocquardAS2016}, Mocquard, Anceaume, and Sericola show how to remove this assumption and make the protocol uniform. Their protocol computes the ratio of $\A$ agents with respect to $n$ within a multiplicative factor error $(1 + \epsilon)$ of the true proportion for any precision $\epsilon>0$ using $2\bceil{\frac{3}{4\epsilon}}+1$ states.

\section{Exact population size counting}
\label{sec:exact-counting}

In the exact size counting problem, the agents aim to compute their population size $n$.
In some 
protocols, to reduce the space complexity, the agents report 
their estimate of $n$
as a function of their internal fields~\cite{DEMST18, BerenbrinkKR19} rather than storing the population size explicitly in their memory. This trick helps the agent to describe numbers that exceeds their memory limit. For example, the agents might store $\mathtt{a} = \floor{\log n}$ but set their output as $2^{\mathtt{a}}$ without explicitly computing the value of $2^{\mathtt{a}}$ to keep their memory usage $\Theta(\log n)$ instead of using linear states. 

\subsection{{Na\"{i}ve} slow protocol}\label{subsec:exact-slow}
A {na\"{i}ve} protocol can count the number of agents in a population using a modified version of the slow leader election protocol. 
All of the agents start in the $\actve$ state holding $1$ token in their $\countt$ variable. 
For consistency with other counting protocols in this section, we name the leader $\actve$,
retaining the standard
pairwise leader elimination
$(\actve , \actve) \to (\actve, \Not\ \actve)$.  
We also change the leader election protocol so that the final remaining $\actve$ agent accumulates all the $n$ tokens.
The rules of this protocol preserve the total sum of the active tokens. When two active agents interact one of them becomes inactive, and both change their $\countt$ value to the sum of their accumulated tokens. 
Initially, the agents start with $n$ scattered tokens and eventually, there will remain one $\actve$ agent having all the tokens. 
Protocol~\ref{protocol:slow-exact-counting} describes how the agents update their state at each interaction. 



\begin{algorithm}[ht]
	\floatname{algorithm}{Protocol}
	\caption{$\exactslow(\rec, \sen)$\\
	    \textbf{Initialization: }\\
	    $\agent.\countt \gets 1$; \quad
	    $\agent.\actve \gets \True$
	}
	\label{protocol:slow-exact-counting}
	\begin{algorithmic}[1]
	    \If{$\rec.\actve = \True \And \sen.\actve = \True$}
	        \State {$\rec.\countt, \sen.\countt \gets \rec.\countt + \sen.\countt$}
	        \State {$\sen.\actve \gets \False$}
	    \EndIf
		\If{$\rec.\actve= \False \And \sen.\actve = \False$}
	        \State {$\rec.\countt, \sen.\countt \gets \text{max}(\rec.\countt , \sen.\countt)$}
	    \EndIf
	\end{algorithmic}
\end{algorithm}



The transitions of Protocol~\ref{protocol:slow-exact-counting} require $O(n)$ time to converge to the exact population size $n$.\footnote{
    This is a standard analysis in population protocols; for instance, see~\cite[Section 6]{AADFP06}. One way to see it requires $\Omega(n)$ time is to observe that once exactly two agents have $\actve = \True$, since there are $\binom{n}{2} = \Theta(n^2)$ total pairs of agents, it takes expected $\Theta(n^2)$ interactions, i.e., $\Theta(n)$ expected parallel time, for the two $\actve$ agents to interact and reduce their count to one.
}
$\Omega(n)$ is a clear lower bound on the number of states 
needed for any protocol 
that requires agents to store the value $n$, 
since $\ceil{\log n}$ bits are required merely to write the number $n$. Protocol~\ref{protocol:slow-exact-counting} solves this problem using $2n$ states.

\subsection{Fast exact counting with polynomial states}\label{subsec:doty-exact}
Doty, Eftekhari, Michail, Spirakis, and Theofilatos~\cite{DEMST18} devised the first sublinear time protocol for the exact population size counting problem. 
Although their protocol heavily relies on the idea of the averaging protocol of~\cite{MocquardAABS2015} (explained in Section~\ref{sec:prereq-mocquard-average}), they managed to eliminate the ``advance knowledge of $n$'' assumption. Their protocol achieves uniformity by putting together one phase of leader election and approximate counting before the averaging phase. 
Additionally, the averaging part of~\cite{DEMST18} is slightly different from the protocol of~\cite{MocquardAABS2015} by fixing the number of agents in groups $\A$ and $\B$ and changing their initial values.
Recall that in Protocol~\ref{protocol:avg-one-way}, the $\A$, $\B$ agents start with $+m$, $-m$ respectively and they converge to $\frac{(n_a - n_b)m}{n}$.
In the protocol of~\cite{DEMST18}, the agents start the averaging phase in a very special case of one $\A$ agent with $+m$ tokens and $n-1$ of $\B$ agents with $0$ tokens. Following the rules of Protocol~\ref{protocol:avg-one-way}, the agents' $\ave$ value will converge to $\approx m/n$. The authors proved with any values of $m \geq 3n^3$, $\floor{\frac{\ave}{m}+ \frac{1}{2}}$ will be equal to $n$ after $O(\log n)$ time. Berenbrink, Kaaser, and Radzik improved this result showing that the correctness of the above statement holds for smaller values of $m$ as long as $m \geq 4n^2$~\cite[Lemma 4.2]{BerenbrinkKR19}, which implies a better space complexity since we can initialize the agents with smaller values of $m$ in the averaging protocol, preserving the correctness of the output.

It remains to show how to initialize the population with the above requirement: 
having one agent in group $\A$, called leader, with $m \geq 3n^3$ tokens.
To achieve this, the 
protocol of~\cite{DEMST18} begins by assigning unique $\codename$s (binary strings), of the same length, to all the agents.
All agents start with the empty string $\varepsilon$ as their $\codename$.
New $\codename$s are generated dynamically whenever two agents with the same $\codename$ $= x_1x_2 \ldots x_l$ of length $l$ interact; each decides a new $\codename$ of length $2l$ by appending $l$ more random bits to their $\codename$. 
Also, if one agent has a longer $\codename$ than its partner, the latter appends random bits until their $\codename$ lengths are the same.
Once the agents have unique $\codename$s of length $l^*$, 
it is shown that
$\log n \leq l^* \leq 3 \log n$ 
holds with high probability. 
This provides each agent with a polynomial-factor approximation of $n$ that is in $[n, n^3]$.  
The leader election protocol of~\cite{DEMST18} is as simple as a pairwise comparison of $\codename$s, adopting the lexicographically largest $\codename$ as the leader.
Once there is a unique leader and an estimate of $n$, 
the agents start the averaging subprotocol.
This protocol as designed is 
stabilizing (correct with probability $1$) and converges to the correct value of $n$ after $\Theta(\log n \log \log n )$ time both w.h.p. (probability at least $1-\frac{O(\log \log n)}{n}$) and in expectation. The error of having multiple leaders always will be detected (since all agents eventually have unique $\codename$s) and the agents replace their $\outtput$ value resulting an always correct protocol. 

\subsection{Fast exact counting with linear states}\label{subsec:berenbrink-exact}
Berenbrink, Kaaser, and Radzik~\cite{BerenbrinkKR19} improved the space complexity as well as the time complexity of the exact counting protocol of~\cite{DEMST18}. 
They start their exact counting with a subprotocol of~\cite{GS18} that elects not a single leader but a ``junta'' of $n^\epsilon$ leaders, for $0 \leq \epsilon < 1$.
In the junta election protocol of~\cite{GS18} (see Section~\ref{subsec:junta-election} for more details), each agent computes a $\level$ value, and the maximum $\level$ among the agents is an approximation of $\log \log n$: 
assuming $l^*$ is the maximum level,
$\log \log n -4 \leq l^* \leq \log \log n +8$ with probability at least $1-O(1/n)$~\cite[Lemma 4]{Berenbrink2020majority_tradeoff}~\cite[Theorem 3]{GS18}.
Having a junta of size $n^\epsilon$, opens the possibility of simulating a ``phase clock'' that allows agents to stay synchronized within phases of length $\Theta(\log n)$ for $\poly(n)$ time~\cite{AAE08, GS18, Berenbrink2020majority_tradeoff,alistarh2018advances}.
In addition to the junta, the exact counting protocol of~\cite{BerenbrinkKR19} requires having one unique leader. 

They use the leader election protocol 
from~\cite{Berenbrink2020majority_tradeoff} that uses constant number of phases to elect a leader: in every even phase, each remaining leader generates a sequence of $\Theta(\log n)$ random bits. In the odd phases, they broadcast the 
maximum bitstring by epidemic and if a leader encounters a larger bitstring than its own, it updates its state to follower. This leader election protocol is a generalization of the $O(\log^2 n)$ time protocol described in~\cite{GS18},
which allows remaining leaders to generate and broadcast $1$ random bit in each phase and continues for $O(\log n)$ phases of each $O(\log n)$ time.

In the rest of the protocol, we assume there exists a leader and the agents 
all hold the value $l^*$ computed as described above. Moreover, the agents are synchronized via the junta-driven phase clock that gives them phases of $\Theta(\log n)$ time. Note that, the averaging process explained in Protocol~\ref{protocol:avg-one-way} takes $\Theta(\log n)$ time to almost equally distribute the initialized load.
At this point, it is possible to adopt the technique of~\cite{DEMST18} explained in Section~\ref{subsec:doty-exact} and, using the fact that $2^{2^{l^*+4}} \geq n$, initialize the leader with at least $n^2$ tokens.
However, this approach 
leads to a protocol that uses at least $2^{2^{\log \log n+12}}=2^{4096\log n} = n^{4096}$ states (when $l^* = \log \log n+8$), which is worse than $O(n^{60})$ states of the protocol of~\cite{DEMST18}. 
In~\cite{BerenbrinkKR19}, the authors refine the approximation value of $\log n$
through possibly multiple (constant) phases of $O(\log n)$ time each, that eventually a total of at least $2n$ tokens will be distributed among agents:

The leader initializes the averaging process with
$2^{2^{l^*-8}}$ tokens (note that $2^{2^{l^*-8}} \leq n$) and signals the agents to multiply the total number of tokens by $2^{2^{l^*-8}}$ followed by an averaging phase until the total number of distributed tokens is less than $2n$. 

Specifically, by the end of each averaging phase, if the leader's $\ave$ is less than $4$, all the agents (including leader) multiply their $\ave$ by another $2^{2^{l^*-8}}$ and repeat the averaging process. Note that this will multiply the total number of tokens by $2^{2^{l^*-8}}$. Depending on the precision of $l^*$ for $\log \log n$, this process 
may take multiple (constant) phases of multiplication followed by an averaging phase, and stops once the leader has $\ave \geq 4$. At this point, the leader computes an approximation of $\log n$ (stores in $k$) as a function of $(p_i, l^*, \ave)$ (precisely, set $k = p_i \cdot 2^{l^* - 8 } - \floor{\log(\ave)}$) in which $p_i$ indicates how many times the agents multiplied their $\ave$ value. The authors proved $\log n - 3 \leq k \leq \log n +3$ holds w.h.p.

In the next stage of the protocol, the agents compute the exact value of $n$ using the computed $k$ value as an approximation of $\log n$ via two phases of averaging:

The leader broadcasts $k$, to all agents and initialize a new averaging process with $c \cdot 2^k$ tokens where $c = 2^8$ and is a constant. The agents distribute $c \cdot 2^k$ tokens through the averaging phase and by the end of it, all agents multiply their $\ave$ value by $2^k$ (once) and repeat the averaging. By the end of this phase, a total of at least $n^2$ tokens has been distributed. Thus, the agents can compute the exact value of $n$, similar to~\cite{DEMST18}, as a function of $(c \cdot 2^{2k}, \ave)$.

In contrast to the protocol of~\cite{DEMST18}, where the leader starts with $\poly(n)$ tokens ($n^c$ for $3\leq c\leq 9$), at every stage of the protocol of~\cite{BerenbrinkKR19}, the leader starts with no more than $n$ tokens.
Once the agents have almost equal tokens because of the averaging phase, the entire population multiplies their $\ave$ value (tokens) by another factor of $\approx n$. This trick puts an upper bound of $n$ over the range of possible values of $\ave$ but achieves having a total of $\poly(n)$ tokens among the population. The protocol of~\cite{BerenbrinkKR19} uses $O(n \log n)$ states and converges in $O(\log n)$ time both w.h.p. 
However, this protocol has a small probability of error; i.e., it is not \emph{stable} (See Section~\ref{subsec:defs}).
It is explained in~\cite{BerenbrinkKR19} how to achieve stabilization in $O(\log n)$ time using $O(n \log n \log \log n)$ states with error detection schemes that point agents to switch to the {na\"{i}ve} slow (but stable) Protocol~\ref{subsec:exact-slow} as a backup.

\subsection{Population protocols with constant size messages}\label{subsec:constant-message}
Amir, Aspnes, Doty, Eftekhari, and Severson~\cite{doty2020messagecomplexity} studied the exact counting problem in population protocols with large memories but limited (constant) message size. Considering the exact population size counting problem in this model, the authors of~\cite{doty2020messagecomplexity} proposed a leader-driven protocol to count the exact population size that converges in $O(\log^2 n)$ time using $O(n\log^2 n)$ states with probability at least $1-O(1/n)$.
They also proposed a leaderless protocol that counts the exact number of agents in a population using $O(\log^2 n)$ time and $O(n\ \polylog\ n)$ states with probability at least $1-O(1/n)$. 
They also demonstrated protocols that \emph{approximate} the population size, also using $O(1)$ messages. See Section~\ref{sec:approx-counting} for a definition of approximate population size counting.


The following large-message protocol allows agents to compute $n$:
the leader starts with value $1$, and agents conduct a rational-number variant of the averaging protocol (e.g., 
$1,0 \to \frac{1}{2},\frac{1}{2}$;
\quad 
$\frac{1}{2},0 \to \frac{1}{4},\frac{1}{4}$;
\quad
$\frac{1}{4},\frac{1}{8} \to \frac{3}{16},\frac{3}{16}$)
until all agents hold dyadic values close enough to $\frac{1}{n}$ that they can uniquely identify the size $n$.
The protocol of~\cite{doty2020messagecomplexity} simulates this in $O(\log n)$ phases 
(synchronized via a leader-driven phase clock),
averaging together only 
\emph{constantly} many 
values at a time,
narrowing the interval of values stored internally by agents, until it contains a unique integer reciprocal $\frac{1}{n}$.

\subsection{Self-stabilizing counting}
\label{subsec:self-stabilizing}
So far we have discussed the \emph{initialized} setting, where we assume the protocol is permitted to designate a set of \emph{valid} initial configurations.
In the case of the counting problem, we identify a special state $x_0$, where valid initial configurations have all agents in state $x_0$.
In contrast, in the \emph{self-stabilizing} setting,
once the set of states has been defined by the protocol,
an adversary can initialize the population with an \emph{arbitrary} configuration assigning these states to agents.
This is an extreme form of fault tolerance, modeling errors that can alter states arbitrarily, at any time during the execution of a protocol, requiring the protocol to be able to recover from any number of such transient errors,
by considering the ``initial'' configuration to be the (arbitrary) configuration just after the \emph{last} such transient error.

It is worth observing why counting, as defined previously, is impossible in this strict setting.
Suppose that a population of $n$ agents has stabilized on output $n$. 
Then for any $k < n$, 
in the self-stabilizing setting,
any configuration of a sub-population of $k$ of these agents is a valid starting configuration for population size $k$.
Then this size-$k$ population must eventually change their output from $n$ to $k$.
However, the interactions that achieve this are possible in the size-$k$ sub-population of the original size-$n$ population, contradicting its stability.

To circumvent this impossibility,
protocols for the self-stabilizing counting problem 
have considered adding one exceptional entity, called the \emph{base station}, such that the adversary is not permitted to affect its memory~\cite{AspnesBBS2016, beauquier2015space, BeauquierSS2007, IZUMI2014}. 
Furthermore, only the base station is required to know the count after stabilization; thus it is possible for other agents to have fewer than $n$ states.
In these protocols, the base station stably computes the exact number $n$ of agents in the population, called \emph{counted} agents.
Assuming a known upper bound $P$ on the population size $n$, Beauquier, Clement, Messika, Rosaz, and Rozoy~\cite{BeauquierSS2007} proposed a protocol that solves the exact counting problem using $4P$ states. This result improved by Izumi, Kinpara, Izumi, and Wada~\cite{IZUMI2014} to $2P$ states space per counted agent. In both protocols, the base station assigns unique names to the counted agents. 
Beauquier, Burman, Clavi{\`e}re, and Sohier~\cite{beauquier2015space} proposed a space-optimal protocol that solves the exact counting problem using $1$-bit memory for each counted agent in $O(2^n)$ time. Later on, 
Aspnes, Beauquier, Burman, and Sohier reduced the exponential time complexity to $O(n \log n)$ time which is also proven to be optimal while still using $1$-bit memory for the counted agents~\cite{AspnesBBS2016}.

\section{Approximate population size counting}
\label{sec:approx-counting}

The study of the counting problem is partially motivated by the existence of nonuniform protocols. Most of these nonuniform protocols require not $n$ exactly, but an approximation, e.g., the value $\ceil{\log n}$. 
In the approximate size counting problem, the agents compute an approximation of $n$ , 
e.g., $2^{\ceil{\log n}}$, 
the smallest power of two greater than $n$, 
rather than the exact value $n$. 
This freedom opens room for protocols with exponentially smaller space complexity.

\subsection{Na\"{i}ve slow protocol}\label{subsec:approx-slow}
Recall that Protocol~\ref{protocol:slow-exact-counting} 
solves the exact counting problem via pairwise elimination of active agents, passing all the tokens (where each agent starts with one token) to the remaining active agent.
A simple modification to Protocol~\ref{protocol:slow-exact-counting} can solve the approximate counting
problem using $O(n \log n)$ time~\cite{Berenbrink2020majority_tradeoff, doty2021majority}. 
In the protocol presented next,
token counts are restricted to powers of two, thus using only $\Theta(\log n)$ states. 
All agents start in the $\actve$ state with one token stored in their 
$\exponent$ field (initially set to $0$ representing integer $2^0$). When two $\actve$ agents with the same $\exponent$ value equal to $i$ (integer value of $2^i$) interact, one of them becomes $\Not\ \actve$, and both update their $\exponent$ to $i + 1$ (integer value of $2^{i+1}$). Additionally, all $\Not\ \actve$ agents help propagating the maximum value of $\exponent$ they have seen (described in Protocol~\ref{protocol:slow-approx-counting}). 

\begin{algorithm}[ht]
	\floatname{algorithm}{Protocol}
	\caption{$\approxslow(\rec, \sen)$\\
	    \textbf{Initialization: }\\
	    $\agent.\exponent \gets 0$;\quad
	    $\agent.\actve \gets \True$
	}
	\label{protocol:slow-approx-counting}
	\begin{algorithmic}[1]
	    \If{$(\rec.\actve \And \sen.\actve) \And (\rec.\exponent = \sen. \exponent)$}
	        \State {$\rec.\exponent, \sen.\exponent \gets \rec.\exponent + 1$}
	        \State {$\sen.\actve \gets \False$}
	    \EndIf
		\If{$\rec.\actve = \False \And \sen.\actve = \False$}
	        \State {$\rec.\exponent, \sen.\exponent \gets \text{max}(\rec.\exponent , \sen.\exponent)$}
	    \EndIf
	\end{algorithmic}
\end{algorithm}


Although Protocol~\ref{protocol:slow-approx-counting} is slow and takes $O(n\log n)$ time, it utilizes almost optimal space complexity. This protocol uses $O(\log n)$ states having $n-O(\log n)$ agents store $\floor{\log n}$; however, requiring all agents to store $\floor{\log n}$ results in a $O(\log ^2 n)$ state protocol~\footnote{
For $n = 2^k, k \in \N$, the population converges to having one unique $\actve$ agent, and all $\Not\ \actve$ agents will store the floor of $\log n$. 
For other values of $n \neq 2^k, k\in \N$, the population converges to $O(\log n)$ $\actve$ agents each having a different value of $\{0, \ldots, \floor{\log n} \}$ that results in all null interactions. Note that the interaction between ($\actve$, $2^3$), ($\actve$, $2^5$), concludes with both agents having the same states.

To be concrete, exactly $b_1$ $\actve$ agents will remain, such that $b_1$ is the number of $1$s in the binary expansion of $n$. Each of the $b_1$ $\actve$ agents hold one of the values $i_1, \ldots, i_{b_1}$ for all the indices that have $1$ in the binary expansion of $n$. Thus, to enforce ``all'' agents (both $\actve$ and $\Not \actve$) report the value of $\log n$, the protocol needs at most $O(\log^2 n)$ states per agent. 
}. Note that $\ceil{\log \log n}$ bits (equivalently $\Theta(\log n)$ states) are needed to write the number $\floor{\log n}$ or $\ceil{\log n}$ for any protocol that reports an estimation of $\log n$ as its output. 

In the following, we overview the fast protocols that considered the approximate counting problem. Commonly, the output of these protocols is an approximation of $\log n$. 

\subsection{Maximum of n geometric random variables} \label{subsec:n-geometrics}


Assuming a randomized protocol, i.e., agents have access to independent, unbiased random bits, there is a simple method for obtaining a constant-factor approximation of $\log n$, i.e, a polynomial factor approximation of $n$.
Recall that a \emph{$\frac{1}{2}$-geometric} random variable is the number of flips of a fair coin until the first heads.
It is known that the maximum of $n$ independent $\frac{1}{2}$-geometric random variables is in the interval $[\log n - \log \ln n, 2 \log n]$ with probability at least $1 - O(1/n)$~\cite{doty2018efficient,eisenberg2008expectation}.
Each agent flips a fair coin on each interaction, incrementing a counter until the first heads,\footnote{
    In more powerful variants of the model, each agent runs a randomized Turing machine~\cite{CMNPS11, DEMST18, doty2018efficient}.
    In this case the $\frac{1}{2}$-geometric random variable can be generated in one step.} 
and then moves to a ``propagate the maximum'' stage where the maximum counter value obtained by any agent is spread by epidemic throughout the population, i.e., $i,j \to i,i$ if $i>j$.

\subsubsection{Synthetic coins}
Since it may be desirable to use a deterministic transition function, some work has been done on techniques for simulating randomized transitions with a deterministic transition function.
Alistarh, Aspnes, Eisenstat, Gelashvili, Rivest~\cite{alistarh2017time}
proposed a general technique, known as \emph{synthetic coins}, that synthesizes  ``almost'' independent and 
unbiased random coin flips in a deterministic protocol, ``extracting'' randomness from the random scheduler.
Each agent uses this synthetic coin technique to simulate generating a $\frac{1}{2}$-geometric random variable $G_i$.
Their protocol provides an approximation of $\log n$ in the interval $[1/2 \log n, 9\log n]$ with probability at least $1-O(1)/n^3$, 
i.e., worse bounds than obtained with independent, unbiased coin flips, but still within a constant factor of $\log n$.

Recently, Sudo, Ooshita, Izumi, Kakugawa, and Masuzawa~\cite{sudo2018logarithmic} proposed an improved implementation of synthetic coins: independent and unbiased coins (as with~\cite{alistarh2017time}, using only symmetric transitions).
The method of Sudo et al.~\cite{sudo2018logarithmic} works as follows for any protocol where ``population splitting'' can be used. 
(See~\cite[Section 4.3]{alistarh2018advances}.)
Create a subpopulation of ``coin'' agents whose only job is to provide random bits to the remaining ``main'' agents.
Main agents build up a list of random bit values to use in the main algorithm, 
which they obtain when interacting with coin agents.
Coin agents start (after first being assigned to the coin subpopulation)
in state $J$, with the following transitions:
$J,J \to K,K$;
$K,K \to J,J$;
$J,K \to C_0,C_1$.
When a main agent interacts with $C_b$, it appends bit $b$ to its list of random bits.
Since the above transitions ensure that there are exactly the same number of $C_0$ and $C_1$ agents at any time, the bits are unbiased.
Since the scheduler ensures that, conditioned on an interaction being between a main and a coin agent, 
the choice of coin agent is independent of other main-coin interactions,
the bit values built up in main agents are independent.\footnote{
    The only difference with a truly randomized protocol is that main agents may have to wait to build up random bits before being allowed to do a randomized transition with another agent.
    This does introduce some dependence in the main protocol, which means this is not a fully black-box technique for replacing a randomized protocol with a deterministic protocol.
}

\subsection{Arbitrary biased coins}\label{subsec:junta-election}
One can approximate $\log \log n$ with access to random bits with arbitrary bias. 
We explained above how to approximate $\log n$ with a series of $\frac{1}{2}$-biased coins. 
Consider instead a special coin whose initial bias (probability of tail, i.e., continuing to flip) of $\frac{1}{2}$
is squared 
after each coin flip.
In other words, the bias is $\frac{1}{2}$ for the first flip, 
$\frac{1}{4}$ for the second flip, 
$\frac{1}{16}$ for the third flip, etc. Similarly to the previous protocol, let each agent independently flip this special coin until a head appears and store the number of consecutive tails. 
In this process, 
the fraction of agents who get a tail and continue flipping is approximately squared after each flip, so 
the maximum stored value among $n$ agents is an approximation of $\log \log n$.
The junta election protocol of~\cite{GS18} (also explained in~\cite{Berenbrink2020majority_tradeoff}) simulates this process without using any coin flips. 
We describe the modified version of protocol~\cite{GS18} for simplicity~\cite{Berenbrink2020majority_tradeoff}. In this protocol 
the agents store their current coin number using a $\level$ variable. Initially, all agents start in state $(\level = 0, \actve = \True)$, and eventually, all will set their $\actve$ to $\False$. Agents increase their $\level$ value via the \emph{asymmetric} transition rules indicated in Protocol~\ref{protocol:junta-counting}.

\begin{algorithm}[ht]
	\floatname{algorithm}{Protocol}
	\caption{$\approxjunta(\rec, \sen)$\\
	    \textbf{Initialization: }\\
	    $\agent.\level \gets 0$; \quad 
	    $\agent.\actve \gets \True$
	}
	\label{protocol:junta-counting}
	\begin{algorithmic}[1]
	    \If{$\rec.\actve \And \rec.\level = 0$}
	        \State {$\rec.\level \gets 1$}
	        \Comment{happens at the first interaction}
	    \EndIf
	    \If{$\sen.\actve \And \sen.\level = 0$}
	        \State {$\sen.\actve \gets \False$}
	        \Comment{happens at the first interaction}
		\ElsIf{$\rec.\actve \And \rec.\level> 0$}
    		\If{$\rec.\level < \sen.\level$}
    		    \State {$\rec.\level \gets \rec.\level+1$}
    		\ElsIf{$\rec.\level \geq \sen.\level$}
    		    \State {$\rec.\actve \gets \False$}
    		\EndIf
	    \EndIf
	\end{algorithmic}
\end{algorithm}



The combination of the first two if statements in~\ref{protocol:junta-counting} acts similarly to the $\frac{1}{2}$-bias coin. About $n/2$ agents participate in their first interaction as receiver and increase their $\level$ by $1$. 
Intuitively, in Protocol~\ref{protocol:junta-counting},
with $\alpha n$ agents (such that $0 < \alpha < 1$) having $\level \geq i \geq 1$, there will be about $\alpha^2 n$ with $\level \geq i+1$.


It is proven that the maximum $\level$ value in the population ($l^*$) is an additive approximation of $\log \log n$.
More precisely, $\log \log n -4 \leq l^* \leq \log \log n +8$ with probability at least $1-O(1/n)$~\cite{GS18, Berenbrink2020majority_tradeoff}.
This yields a multiplicative factor approximation of $\log n$; see Table~\ref{table:approx-counting-summary}.

The above protocol is also a so-called \emph{junta election} protocol:
the number of agents who obtain the maximum $\level$ is $O(\sqrt{n \log n})$ with high probability.
These agents can be used, for example to create a ``junta-driven phase clock''\cite{GS18}, useful for synchronization.

\subsection{Fast protocol with additive error}\label{subsec:doty-approx}
Doty and Eftekhari~\cite{doty2018efficient} presented a protocol that improves 
the approximation factor of the protocol of~\cite{alistarh2017time}, which approximates $\log n$ using maximum of $n$ geometric random variables, from a multiplicative to an additive factor approximation. Their protocol converges to an estimation of $\log n$ in the interval $[\log n -5.7 , \log n +5.7]$ after $O(\log^2 n)$ time using $O(\log^4 n)$ number of states per agent. They extend one round of taking the maximum of $n$ geometric random variables of~\cite{alistarh2017time} to $K$ rounds of taking maximums and computing their average as an approximation of $\log n$. 
Doty and Eftekhari~\cite{doty2018efficient} proved that the computed average is within $O(1)$ of $\log n$ with $K = \Omega(\log n)$. In their protocol, the agents agree on $K$ by taking the maximum of $n$ independent geometric random variables. For the rest of the protocol, the agents simulate a uniform variation of the leaderless phase clock introduced in~\cite{alistarh2018space} to synchronize the agents for $K = O(\log n)$ rounds. In a leaderless phase clock, all agents individually\footnote{The leaderless phase clock of the protocol~\cite{doty2018efficient} allows agents to increment their counts at every interactions: $c_i, c_j \to c_{i+1}, c_{j+1}$. In contrast, in a ``leaderless phase clock with power of two choices''~\cite{alistarh2018space, bennun20majority} if two agents with counts $i$ and $j$ interact, only the agent with smaller count value increments:
$c_i, c_j \to c_{i+1}, c_j$ for $c_i \leq c_j$. The latter phase clock keeps agents' count values tightly close to each other.} count their number of interactions and compare it with a threshold value $\Theta(\log n)$. If agents' counts reach the threshold, they simply move to the next round of the protocol and set their count value to zero. 
In each round of the protocol, the agents generate one new geometric random variable, propagate it, and store the maximum. After $K$ rounds, the agents learn $K$ values, each is a maximum of $n$ independent geometric random variables, in sequence and take their sum. In round $K+1$, the agent divides the sum by $K$ and stores the result as their output. 

The protocols we have discussed so far for approximating the population size are leaderless. They are not dependent on the existence of a unique leader. In a leaderless protocol, all agents are initially equivalent, and there is no distinguished leader. 

\subsection{Leader-driven, epidemic-based protocol}\label{subsec:michail-approx}
A leader-driven protocol for approximating the population size was introduced in~\cite{michail2018fast}. The basic idea of their protocol relies on the completion time of an epidemic process. Specifically, the leader triggers an epidemic process (infecting exactly one agent $L, Q \to L^*, A$) and keeps track of the number of infected ($c_a$) and uninfected ($c_q$) agents without infecting more agents. The followers (initialized in state $Q$) participates in this protocol via the one-way epidemic rule ($A, Q \to A, A$).
As soon as the number of infected and uninfected agents becomes equal, the leader terminates the protocol and reports the approximation $n' = 2^{c_a + 1}$. In~\cite{michail2018fast} it was shown that $c_a \leq 2\log n$ with high probability.\footnote{Theorem 1 of~\cite{michail2018fast} states that $\log n \leq c_a$ with high probability. However, this does not appear to hold in simulation. It seems likely that a bound of $c \cdot \log n$ can be proven for some $c>0$ based on known results lower bounding times for epidemics to spread~\cite{mocquard2016epidemic}.}
This protocol approximates the population size using $\Theta(\log n)$ states for leader while the followers use constant states.

\subsection{Discrete averaging with powers of two}\label{subsec:berenbrink-approx}
Berenbrink, Kaaser, and Radzik~\cite{BerenbrinkKR19} introduced a new averaging protocol via modifying the rules of Protocol~\ref{protocol:avg-one-way}. Recall that Protocol~\ref{protocol:avg-one-way} works by pairwise averaging of nonnegative integer values held by each agent; the modified rules restrict the agents to use numbers that are a perfect power of two. 
The authors carefully developed the protocol such that the new rules of the protocol still preserve the total sum among the agents.
In this variation of the averaging protocol, shown in Protocol~\ref{protocol:avg-power-two}, the agents can store either a perfect power of two or zero. 
The constraint helps to reduce the space complexity via representing an integer $2^x$ with $x$. To show the exact value of $0$, the agents use $-1$. 
Using a similar approach to~\cite{DEMST18, BerenbrinkKR19} for the exact counting problem, a leader starts an averaging process with a large (with respect to $n$) positive value, and all the followers start with zero ($\ave = -1$). 

\begin{algorithm}[ht]
	\floatname{algorithm}{Protocol}
	\caption{$\powertwoaveraging(\rec, \sen)$\\
	    \textbf{Initialization: }\\
	    if $\agent.\mathtt{leaderBit} =\True$: $\agent.\ave \gets m$\\
	    if $\agent.\mathtt{leaderBit} = \False$: $\agent.\ave \gets -1$
	}
	\label{protocol:avg-power-two}
	\begin{algorithmic}[1]
	    \If{$\rec.\ave = -1 \And \sen.\ave > 0$}
    	    \State {$\rec.\ave, \sen.\ave \gets
    		\sen.\ave -1$
    		}
    	\ElsIf{$\sen.\ave = -1 \And \rec.\ave > 0$}
        	\State {$\rec.\ave, \sen.\ave \gets
        		\rec.\ave -1$
        	}
	    \EndIf
	\end{algorithmic}
\end{algorithm}

Utilizing the restricted version of the discrete average process, they proposed an approximate counting protocol that outputs the value $\floor{\log n}$ or $\ceil{\log n}$ with high probability, using $O(\log^2 n)$ time and $O(\log n \log \log n)$ states. To stably solve the approximate counting problem, they used multiple always correct error detection schemes that point the population to the slow token-passing Protocol~\ref{protocol:slow-approx-counting} if an error occurs. 
With an overhead of $O(\log n)$ states, their protocol stabilizes to $\floor{\log n}$ or $\ceil{\log n}$ using $O(\log^2 n \log \log n)$ states after $O(\log^2 n)$ time. 

For the fast computation of $\log n$, similar to the exact counting protocol of~\cite{BerenbrinkKR19} described in Section~\ref{sec:exact-counting}, all the agents simulate the junta election protocol of~\cite{GS18} at the very beginning to (1) simulate a junta driven phase clock and achieve synchronization and (2) elect a unique leader for the discrete averaging Protocol~\ref{protocol:avg-power-two}. Once the population elected its leader, the leader performs a linear search, starts $0$, to find $\floor{\log n}$ or $\ceil{\log n}$. 
At the beginning of round $i$, the agents reset their $\ave$ back to $-1$ and follow the rules of Protocol~\ref{protocol:avg-power-two} while the leader starts with $\ave = i$ (injecting $2^i$ tokens to the population).
At the end of round $i$, if some agents hold an average value greater than zero ($\ave \geq 1$), the leader will stop the search and broadcast the value $i$ as an approximation of $\log n$. The author proved that w.h.p.~the leader stops the search after $\floor{\log n}$ or $\ceil{\log n}$ rounds. 

\subsection{Regulating size in the presence of an adversary}\label{subsec:goldwasser-size-stability}
Goldwasser, Ostrovsky, Scafuro, and Sealfon~\cite{goldwasser2018population} studied a variation of population protocols,
which allows agents individually to decide to replicate or self-destruct, changing the population size in response to an adversary who can add or remove arbitrary number of agents. 
They proposed a protocol that can approximately maintain a target population size 
(both the initial population size and the target are assumed to be known to each agent) despite this adversary. 


However, they use a \emph{synchronous} variation of population protocols: in one round of the computation, a constant fraction of agents interact (at most once) via a random matching of size $k=O(n)$. 
Observe that, unlike the asynchronous scheduler, the synchronous scheduler prevents any agent having multiple interactions per $k$ total interactions.
Despite this difference in definitions, it is conceivable that techniques used in the analysis of~\cite{goldwasser2018population} could be applied to the standard population protocol model.
It is also noteworthy that their model of agents being created and destroyed is expressible in the model of chemical reaction networks~\cite{SolCooWinBru08}, of which population protocols are a special case.

\section{Tools for making nonuniform protocols uniform}
\label{sec:tools}


Part of the practical motivation behind the study of the counting problem comes from the existence of nonuniform protocols and a desire to create uniform variants of them.
Since most nonuniform protocols require advance knowledge of $\log n$,
the basic technique for making such a protocol uniform is to first compute an estimate of $\log n$ using a protocol from Section~\ref{sec:approx-counting}, then to compose this with the existing nonuniform protocol,
replacing its estimate of $\log n$ with this computed value.
We break this section into two pieces. 
First, we recall a few size approximation protocols from Section~\ref{sec:approx-counting}, 
comparing them in accuracy, space, and simplicity 
with specific suggestions for how to account for these properties in choosing one to be composed with a nonuniform protocol.
In the next part, assuming we have a protocol that computes an estimate of the population size, we show how to compose it with a nonuniform protocol.

\subsection{Comparison of approximate counting protocols}

In this part, we compare techniques that solve the approximate counting problem. 
See Table~\ref{table:approx-counting-summary} for a detailed comparison.
For most nonuniform protocols (e.g., a leaderless phase clock~\cite{alistarh2018space}), a value that is $\Theta(\log n)$ suffices. 
Thus we lead the discussion with protocols that approximate $\log n$ within a multiplicative factor error. 
Although additive approximate error is often unnecessary, in some circumstances,
one may require the estimate be to exactly $\floor{\log n}$ or $\ceil{\log n}$.
For example, the uniform majority protocol of~\cite{doty2021majority} requires the estimate to be at least $\ceil{\log n}$ with probability 1.

\begin{description}
\item[Simple $2$-approximation] [Section~\ref{subsec:n-geometrics}]
Taking the maximum of $n$ geometric random variables provides a $2$-factor approximation of $\log n$~\cite{alistarh2017time, eisenberg2008expectation, doty2018efficient} with probability at least $1-O(1/n)$,\footnote{
    In fact the lower bound is stronger: the maximum is between $\log_2 n  - \log_2 (n) \cdot \ln n$ and $2 \cdot \log_2 n$ with probability at least $1 - O(1/n)$; see~\cite[Lemma 3.8]{doty2018efficient}.
}
and takes $O(\log n)$ time to converge. Although this approach is very simple and straightforward for composition, the space complexity of the protocol is not bounded with probability $1$, and the agents use $O(\log n)$ states w.h.p.

\item[Minimal space overhead] [Section~\ref{subsec:junta-election}] 
Recall that the maximum level $l^*$ in the junta election protocol is $\floor{\log \log n} - 3 \leq l^* \leq \log \log n +4(a+1)$ with probability at least $1-1/n^a$~\cite{Berenbrink2020majority_tradeoff, GS18}. Despite the large multiplicative approximation factor for computing $\ceil{\log n}$, the junta election protocol~\cite{GS18} imposes minimal, $O(\log \log n)$, space overhead and converges in $O(\log n)$ time. Moreover, the protocol provides a junta of $n^\epsilon$ for $0\leq \epsilon <1$ leaders that can simulate a ``junta-driven phase clock'' to synchronize agents in phases of length $\Theta(\log n)$ time~\cite{GS18}. See~\cite{alistarh2018advances, AAE08} for more details about the phase clock.

\item[Maximizing accuracy, always correct] 
Two protocols from sections~\ref{subsec:approx-slow}, \ref{subsec:berenbrink-approx} compute $\floor{\log n}$ (or $\ceil{\log n}$). 
Both protocols provide probability-$1$ correctness using $O(n\log n)$ and $O(\log^2 n)$ time respectively.
The former is much simpler and is used as a ``slow backup'' subroutine in the $O(\log n)$ time protocol of~\cite{doty2021majority}.
Although it is much slower than $O(\log n)$ time,
since it is needed only with low probability,
it contributes negligibly to the expected time.
\end{description}

\subsection{Composition of an uniform counting protocol with a nonuniform protocol}
Most of the time, we can construct a uniform protocol from a nonuniform protocol through composition with a uniform approximate counting protocol. Even though we are unaware of any black-box theorem that proves the correctness of the restarting technique under any circumstance, the authors of~\cite{DEMST18, GS18, doty2018efficient, doty2021majority, BerenbrinkKR19} used the procedure discussed below and proved it correct with an ad-hoc analysis. 
We explain in a general way how to use these approximate counting protocols to make a nonuniform protocol uniform in the next part.

Note that all of the approximate counting protocols mentioned in Section~\ref{sec:approx-counting}, except approximating with a leader explained in Section~\ref{subsec:michail-approx}, are not terminating. In other words, the agents are not aware of the completion of the protocol. 
Although termination is impossible in the uniform model of population protocols~\cite{doty2018efficient}, we can try composing two protocols without using termination of the upstream ones. 
For a concrete discussion, consider protocols \upstream\ and \downstream\ such that \upstream\ is a uniform approximate counting and \downstream\ is a general nonuniform protocol;
\upstream\ is the \emph{upstream} protocol whose output is given as input to the \emph{downstream} protocol \downstream. 
To construct a uniform protocol, we summarize a simple restarting technique that has been used widely~\cite{DEMST18, GS18, BerenbrinkKR19} to compose protocols \upstream\ and \downstream. In this technique, we run both protocols in parallel in the population. If the $\countt$ fields of the protocol \upstream\ in the agents' memory change, then a signal will be propagated (by epidemic) through the population to notify all agents with the updated $\countt$. This signal will stop the protocol \downstream\ (or parts of protocol \downstream\ that are dependent on the population size) and reinitialize it with the updated $\countt$, which eventually will be an approximation of $\log n$. 

Despite the difference between the initial configuration of a nonuniform protocol and the one after the restart signal (agents do not have the same state), any agents who participate in the last execution of protocol \downstream\ restarts their memory (related fields concerning protocol \downstream) to the initial values. Thus, a high probability correct counting scheme can also guarantee the correctness of the downstream protocol \downstream\ w.h.p.

\section{Conclusion and open questions}
\label{sec:conclusion}
In this paper, we gave a brief description of existing protocols that exactly or approximately compute the population size $n$. 
We also discussed a technique for converting nonuniform protocols (those that assume agents are initialized with an approximate estimate of $n$) to uniform protocols,
by composing size approximation with the nonuniform protocol.

While our focus in this paper is the size counting problem, the mentioned protocols demonstrate general techniques that can help solve other problems and design new protocols. Alistarh and Gelashvilli~\cite{alistarh2018advances} mentioned different ideas such as the space multiplexing used in \cite{doty2018efficient, BerenbrinkKR19, Berenbrink2020majority_tradeoff,doty2021majority} and the junta-driven phase clock~\cite{GS18} as available building blocks to design new protocols. 
We also summarized two variations of the discrete averaging technique introduced in~\cite{alistarh2015majority}, 
tightly analyzed in~\cite{MocquardAABS2015, mocquard2019tight,BerenbrinkKR19, berenbrink2019loadbalancing, berenbrink2019tight}, 
that has since been widely deployed in other protocols to solve the counting problem~\cite{DEMST18, BerenbrinkKR19, MocquardAABS2015, MocquardAS2016} and the exact majority problem~\cite{doty2021majority, MocquardAS2016, alistarh2015majority}. 
(See Protocols~\ref{protocol:avg-one-way} and~\ref{protocol:avg-power-two}.)
\paragraph{Open questions with composition of two uniform protocols}
We discussed how to make a nonuniform protocol uniform through composition with a uniform counting protocol that allows the nonuniform protocol to use the output of the counting protocol. 
Generally, in a composition of a uniform protocol \upstream\ with a protocol \downstream, protocol \downstream\ might get restarted repeatedly. In each restart, the agents propagate a new signal with updated information about the size. The counting protocol might even generate a new restart signal before the previous signal hits all the agents. Having this in mind, if a protocol uses duplicate restart signals, restarted and deprecated agents could become indistinguishable. For example, using restart signals of constant size might create inconsistency in the population. 

Unique (and perhaps monotonically increasing) restarting signals guarantees the correctness of the downstream protocol. Since eventually, all agents agree on the last (largest) restart signal and restart protocol \downstream\ for the final execution. 
Even assuming a monotone increasing restart signal might change some probability bounds on the convergence time of protocols. The current literature lacks a general-purpose theorem that proves under what conditions of a downstream protocol the restarting technique works.

\paragraph{Collective output representation of the population size}
Moreover, all the counting protocols summarized in this paper require all agents eventually to represent the computed count. 
If agents are required to store the value $n$, then there is clearly a linear-state lower bound, since $\log n$ bits are required merely to write $n$.
However, what if no agent individually stores all of $n$?
Consider instead a \emph{collective} representation of the population size,
where some agents each store (for example) one bit of $n$,
as well as the significance of the bit.

With this trick, the lower bound does not apply anymore. There might exist a protocol that solves the exact counting with $o(n)$ states. 
However, readout could be more difficult, since we cannot simply look at the memory of a single agent and read the population size; instead, we must sample a small subset of the population. 
For example, composing size computation with another protocol would be less straightforward since the agents who are computing the downstream protocol would not at any point have access to all the bits of $n$.
One could imagine a protocol that spreads the output to the whole population almost equally: for example, having $O(n/\log n)$ agents responsible for each index of the binary expansion of $n$. With this trick, the output will be present dense enough among the population. Thus, a random sample of polylogarithmic agents would have enough information to reconstruct the value of $n$.

The $O(n \log n)$ time slow size approximation protocol (Protocol~\ref{protocol:slow-approx-counting}) collectively represents $n$:
each remaining active agent holds a value $k$ such that there is a 1 at significance $k$ in the binary expansion of $n$.
Is there a sublinear-time, sublinear-state protocol, so that the agents report the population size via this collective representation? 
A valid solution to the exact counting problem with a collective output also solves the \emph{parity} problem: 
compute the least significant bit of $n$.


\bibliography{my-library}
\bibliographystyle{plain}

\end{document}